\documentclass[aps,showpacs,prl,toolkits,twocolumn]{revtex4}
\usepackage{epsfig,amssymb,amsfonts,amsmath,pstricks,psfrag,graphicx}

\makeatletter
\renewcommand{\@makefntext}[1]{\parindent=1em\noindent\hbox to 1.8em
{\hss$^{\@thefnmark}$}#1}
\renewcommand{\@footnotemark}{\hbox{\mathsurround=0pt$^{\@thefnmark}$}}

\makeatother

\begin{document}
\title{A Gap in the Quarkyonic matter}
\author{ L. Ya. Glozman}
\affiliation{Institute for
 Physics, Theoretical Physics branch, University of Graz, Universit\"atsplatz 5,
A-8010 Graz, Austria}

\newcommand{\be}{\begin{equation}}
\newcommand{\bea}{\begin{eqnarray}}
\newcommand{\ee}{\end{equation}}
\newcommand{\eea}{\end{eqnarray}}
\newcommand{\ds}{\displaystyle}
\newcommand{\low}[1]{\raisebox{-1mm}{$#1$}}
\newcommand{\loww}[1]{\raisebox{-1.5mm}{$#1$}}
\newcommand{\lmn}{\mathop{\sim}\limits_{n\gg 1}}
\newcommand{\vpint}{\int\makebox[0mm][r]{\bf --\hspace*{0.13cm}}}
\newcommand{\too}{\mathop{\to}\limits_{N_C\to\infty}}
\newcommand{\vp}{\varphi}
\newcommand{\vx}{{\vec x}}
\newcommand{\vy}{{\vec y}}
\newcommand{\vz}{{\vec z}}
\newcommand{\vk}{{\vec k}}
\newcommand{\vq}{{\vec q}}
\newcommand{\vpp}{{\vec p}}
\newcommand{\vn}{{\vec n}}
\newcommand{\vg}{{\vec \gamma}}

\begin{abstract}
It has recently been suggested  that at a reasonably
large chemical potential a confining quarkyonic matter is formed
that consists of the quark Fermi  sea and confined hadrons
on top of this Fermi sea. We study some  properties of
this matter. It is demonstrated that below the
chiral restoration point  there are gapless excitations of this
matter through excitations of the Goldstone bosons. Above the
chiral restoration point the single quarks are still removed from
the spectrum of excitations and the only possible excitations are
confined color-singlet hadrons with finite mass. Hence there appears
a gap in the excitation spectrum of the quarkyonic matter
that should be crucially important
for its properties  above the chiral restoration
point. This gap is of a new type and is not related with the
condensation of the fermionic system into a quasibosonic system.
It is only due to such properties as confinement and manifest
chiral symmetry at the same time.

\end{abstract}

\maketitle

{\bf 1. Introduction.}
Very recently McLerran and Pisarski have suggested  the unusual
structure of the QCD phase diagram in the large $N_c$ limit \cite{pisarski}.
Previously it was considered natural that at a reasonably large density 
above the critical one and
not so large temperature the QCD matter is a deconfined system of quarks
and gluons, see Fig. 1. In such a situation, however, 
the pressure should be proportional
to $N_c^2$, because it is dominated by the deconfined gluons in the
adjoint representation of $SU(N_c)$. At the very large densities the
perturbative calculation of the pressure is adequate and it produces
a pressure that scales like $N_c$. A resolution of this paradox is
that the system at the reasonably  large densities and low temperatures 
represents
essentially a Fermi sea of valence quarks (baryons are in a strong
overlap and hence it is impossible to decide to which particular baryon a 
given quark belongs) and confined hadrons on top of this sea, see
Fig. 2. Then the
quarks in the Fermi sea produce a pressure $\sim N_c$ while confined
hadrons at the Fermi surface contribute as $\sim 1$. There cannot be 
contribution to the pressure from the deconfined gluons and consequently
the system is manifestly confined. Such a matter was termed as
"quarkyonic".

\begin{figure}
\includegraphics[width=0.5\hsize,clip=]{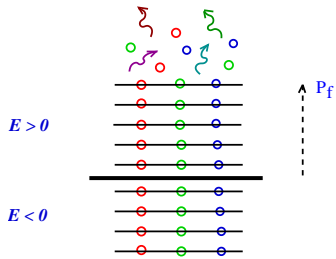}
\caption{A schematic cartoon of the deconfined quark-gluon matter at a finite
chemical potential.}
\end{figure}

\begin{figure}
\includegraphics[width=0.35\hsize,clip=]{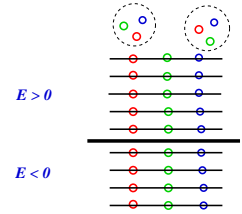}
\caption{A schematic cartoon of the quarkyonic matter at a finite
chemical potential.}
\end{figure}

At the critical chemical potential there should be a chiral restoration
phase transition in this quarkyonic matter. Then on top of the Fermi
sea of quarks there must exist confined but chirally symmetric hadrons.
This  contradicts  to our naive intuition and  to simple models
of confinement and chiral symmetry breaking. It has been demonstrated,
however, that it is not so and actually it is quite natural to expect
confined but chirally symmetric
hadrons on top of the Fermi sea 
above the chiral restoration point \cite{GW}. A possible mechanism for
such hadrons has also been clarified.

Here we would like to address some general and crucially important
properties of the
quarkyonic matter below and above the chiral restoration point,
that have not been  discussed in the papers above.
In the chirally broken phase there
always exists a gapless excitation of the quarkyonic matter, that
is due to excitations of the Goldstone bosons on top of the
Fermi sea. Above the chiral restoration point, however, there is
no a gapless excitation of the quarkyonic matter. The gap is given
by the energy of lowest lying pion-like excitation, that has a {\it
finite} energy which depends on the chemical potential.
Then the properties of the quarkyonic matter are essentially determined
by the presence of the gap. This is quite an unusual situation, because
typically a gap in the excitation spectrum of the Fermi system is
attributed to the condensation of the fermion pairs due to some pairing
interaction between fermions that leads to the quasi bosonic system with
such properties like superconductivity or superfluidity \cite{BCS,B,CS}. 
In our case,
however, a presence of the gap at the chemical potential
above the critical one is not due to condensation of the
Cooper pairs. It has a completely different nature related to such
simultaneous properties of the system as confinement and manifest chiral 
symmetry.
We are not aware of any example of this kind in the usual Fermi systems
and consequently the quarkyonic matter above the chiral restoration
point would represent a Fermi system with  new and yet
unknown properties.

 {\bf 2. Quarkyonic matter below and above the chiral
restoration point.}
The property which we discuss, namely the absence of a gap in the
quarkyonic matter below the critical  chemical potential
and its presence above the critical chemical potential is a 
very general one and is actually related only to confinement
and existence/nonexistence
 of the massless Goldstone excitations at
different chemical potentials. To see it explicitly 
and to understand microscopical reasons one
needs a model that is manifestly chirally symmetric, confining
and provides spontaneous breaking of chiral symmetry. 
Such a model might  not  be exactly equivalent to QCD. However,
 if it incorporates
these properties it provides an insight and illustrates a general
QCD property of the quarkyonic matter in the large $N_c$ limit.

In this context we use the only known exactly solvable confining and
chirally symmetric model in four dimensions that is a generalization
of the 1+1 dimensional 't Hooft model \cite{HOOFT}. 
It is assumed
within this model that there is only a confining Coulomb-like 
linear interquark interaction. Then it manifestly provides spontaneous
breaking of chiral symmetry in the vacuum via the self-energy loops and
there appear Goldstone bosons \cite{Orsay,Adler:1984ri}. 
The complete spectrum of the $\bar q q$ mesons in the vacuum
has been obtained in refs. 
\cite{WG1,WG2} and exhibits a fast  chiral restoration with increasing
$J$, for a review see ref. \cite{GPR}. The meson spectrum at the finite
chemical potential and zero temperature has recently been studied in 
\cite{GW}. It has been
explicitly demonstrated that indeed at the chemical potential below
the critical one the spectrum is similar to the one in the vacuum.
Above the chiral restoration point the quarks are still confined
and the physical spectrum consists of a complete set of exact chiral
multiplets. In the present Letter we will essentially rely on the
results of this paper. We will emphasize a crucially important
point, that was not discussed at all: Above the chiral restoration
point in the confined quarkyonic matter there necessarily appears
a gap in the excitation spectrum. This property is explicitly seen
in the context of this model, and is easily understandable. At the same time
it must be a general property of the quarkyonic matter in the large $N_c$
QCD above the chiral restoration point. 

\begin{figure}
\includegraphics[width=1.0\hsize,clip=]{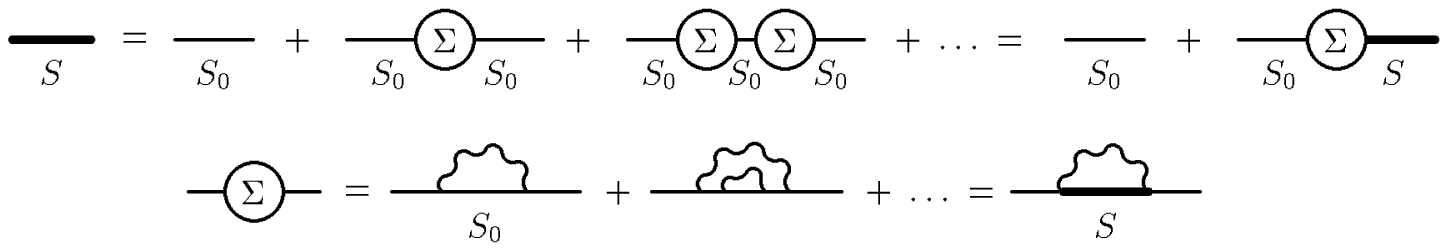}
\caption{Dressed quark Green function and the Schwinger-Dyson equation.}
\end{figure}

We work in the chiral limit and the two flavor 
 version of the model is considered. The global chiral symmetry
of the model is $U(2)_L \times U(2)_R$  because in the large $N_c$
world the axial anomaly is absent.
The model is described
in great detail in references \cite{WG2,GPR} so here we review only
those points that are important for  understanding of the principal
question of the present Letter. 
The only interquark interaction in our case is a linear instantaneous
Lorentz-vector  potential that has a Lorentz
structure of the Coulomb
potential:

\begin{equation} 
K^{ab}_{\mu\nu}(\vec{x}-\vec{y})=g_{\mu 0}g_{\nu 0}
\delta^{ab} V (|\vec{x}-\vec{y}|); ~~~~~
\frac{\lambda^a \lambda^a}{4}V(r) = \sigma r,
\label{KK}
\end{equation}

\noindent
where $a,b$ are color indices. 
Parameterising the self-energy operator in the form

\begin{equation}
\Sigma(\vec p) =A_p +(\vec{\gamma}\cdot\hat{\vec{p}})[B_p-p],
\label{SE} 
\end{equation}

\noindent
where functions $A_p$ and $B_p$ are yet to be found, the
Schwinger-Dyson equation for the self-energy operator in 
the rainbow approximation,which is valid in the large $N_c$ limit
for the instantaneous interaction, see Fig. 3,
is reduced to the nonlinear gap equation for the chiral (Bogoliubov)
angle  $\varphi_p$,
 
 \begin{equation}
 A_p \cos \varphi_p - B_p \sin \varphi_p = 0,
 \label{gap}
 \end{equation}
 
\noindent
where
 
\begin{eqnarray}
A_p & = & \frac{1}{2}\int\frac{d^3k}{(2\pi)^3}V
(\vec{p}-\vec{k})\sin\vp_k,\quad  
\label{AB1} \\
B_p & = & p+\frac{1}{2}\int \frac{d^3k}{(2\pi)^3}\;(\hat{\vec{p}}
\cdot\hat{\vec{k}})V(\vec{p}-\vec{k})\cos\vp_k. 
\label{AB2} 
\end{eqnarray} 

A key point is that the infrared regularization is required for the
Coulomb gauge "gluon" propagator $\sim 1/q^4$. Then all 
observable color-singlet
quantities, like the quark condensate or hadron mass must be independent
on the infrared cut-off parameter in the infrared limit, i.e., when
this parameter approaches 0. On the other hand, the colored quantities,
like the single quark Dirac operator must be infrared divergent, which means
that the single quark is removed from the spectrum and it is confined.
This actually represents a necessary condition for confinement of
quarks. It was demonstrated that indeed in the vacuum all these
requirements are satisfied 
\cite{Orsay,Adler:1984ri,Alkofer:1988tc,WG1,WG2}.

Once the gap equation is solved in the infrared limit one
obtains a quark Green function. Given this quark Green
function we are in a position to solve the Bethe-Salpeter
equation.
The homogeneous Bethe-Salpeter equation for the quark-antiquark
bound states in the rest frame with the instantaneous interaction
is given as 

\begin{eqnarray}
\chi(m,\vpp)&= &- i\int\frac{d^4q}{(2\pi)^4}V(|\vpp-\vq|)\;
\gamma_0 S(q_0+m/2,\vpp-\vq) \nonumber \\
& \times & \chi(m,\vq)S(q_0-m/2,\vpp-\vq)\gamma_0.
\label{GenericSal}
\end{eqnarray}

\noindent
Here $m$ is the meson mass and $\vec p$ is the relative momentum. 
 The infrared
divergence cancels exactly in this equation \cite{WG2}
and it can be solved
either in the infrared limit  or for very small values of the infrared
regulator. Even though the single quark Dirac operator is
divergent in the infrared limit, the color-singlet hadrons are
well defined and finite-energy systems. 
The spectrum of mesons in the vacuum has been calculated in refs. 
\cite{WG1,WG2} and is reviewed in ref. \cite{GPR}. Since the
chiral symmetry is spontaneously broken there are four well defined
Goldstone bosons with zero mass with quantum numbers 
$I=1,0^{-+}$ and $I=0,0^{-+}$. We remind that there are
no vacuum fermion loops in this large $N_c$ model and hence the $U(1)_A$
symmetry is broken only spontaneously.

Assume now that we have a finite chemical potential  at zero
temperature and hence
all levels below the Fermi momentum $p_f$ are occupied.
Consider a "probe quark"
that is brought into the system.
In order to see the properties of this quark
 we have to solve
the gap equation (\ref{gap}) - (\ref{AB2}),
but the integration starts not from $k=0$, but from $k=p_f$,
because all  levels below the Fermi momentum are Pauli blocked.
The points $p < p_f$ are irrelevant for this probe quark
which is always on top of the Fermi sea. The gap equation for
this probe quark has been solved at any finite chemical
potential in ref. \cite{GW} and below we overview those results
that are crucial for our present conclusions.

Once the Fermi momentum $p_f$ is below the critical value $p_f < p_f^{cr}$,
then there is always a nontrivial solution of the
gap equation $\varphi_p \neq 0$ and the chiral symmetry gets dynamically
broken that is manifested in the nonzero dynamical mass of quarks, $M(p)$,
in the nonvanishing quark condensate, as well as in the
presence of the massless Goldstone bosons. However, above
the critical Fermi momentum, $p_f > p_f^{cr}$, there is no nontrivial
solution to the gap equation and the only solution is trivial,
$\varphi_p = 0$, with identically vanishing dynamical mass and
quark condensate. Hence the chiral symmetry gets restored at
$p_f =  p_f^{cr}$ and one obtains the chiral restoration phase
transition at the corresponding chemical potential. Above the chiral
restoration point  the Lorentz scalar self-energy of
quarks identically vanishes, $A_p = 0$. However, the Lorentz spatial-vector 
part $B_p -p$ does not vanish and is in fact the infrared
divergent quantity. Hence the Dirac operator for this single quark
is still infrared divergent.
This implies that even in the chirally restored
phase the single quark  {\it above the Fermi sea}
is confined and cannot be observed\footnote{Below the Fermi surface,
i.e., at $p < p_f$,
there is no the infrared divergence for quarks at all. This is because
the infrared divergence happens at $\vec k = \vec p$, but the integration
in the self-energy integrals $A_p$ and $B_p$ starts from $k = p_f$.}. 
Said differently, this means
that there are no single quark excitations of the quarkyonic matter
even in the chirally restored phase. This property crucially
distinguishes the quarkyonic matter from all usual Fermi systems.

It does not mean, however, that there are no excitations above
the quarkyonic Fermi sea. A color-singlet hadron
constructed out of such probe quarks is a finite and well defined
quantity. The single quark infrared divergence cancels exactly
in the color singlet hadron  and such a hadron is a finite
and well defined quantity even in the chirally restored phase. Of
course, all momenta of quarks inside such a hadron are above the
Fermi momentum. Hence one {\it can excite the system, but the excitation will
consist not of the massless single quarks, but rather of the massive
color singlet hadrons}.
Hadrons with the lowest energy 
 in this case have the
quantum numbers of the  $(1/2,1/2)_a$ and $(1/2,1/2)_b$ chiral multiplets
with $J=0$. In the symmetry broken world these multiplets contain
the Goldstone bosons and the quark condensate.

Above the chiral restoration point their mass is {\it not zero},
see Fig. 4. {\it Then there is a gap in the excitation spectrum of the
quarkyonic matter. This gap can be arbitrary large depending on the chemical
potential.} It is a principal point of the present Letter.
A mechanism for this gap is quite unusual. It
arises not because of the condensation of the Cooper-like pairs in the system.
There is no such a condensation in our case. It arises exclusively
due to the fact that above the chiral restoration point the lowest
 excitation is a confined but chirally symmetric  
 hadron with the non zero mass.

Below the critical Fermi momentum the system is confined and the chiral
symmetry is broken. There are   gapless excitations in the system:
One can always create an arbitrary amount of Goldstone bosons. Above
the chiral restoration point the system is still confined, but the chiral
symmetry is restored. There are neither Goldstone excitations 
nor the single quark excitations in this case.
Then there appears a gap in the excitation
spectrum. This gap will essentially determine properties of the quarkyonic
matter above the chiral restoration point. In particular, a presence
of the gap implies some processes in the system that proceed without
dissipation of  energy. As it is well known, a presence of a gap
is crucial for formation of the superfluid and superconducting systems.
In our case the gap happens, however, without condensation of the
fermionic system into a quasibosonic system. Hence one should expect
that some properties of the quarkyonic matter above the chiral
restoration point should be different as compared to the standard
quasibosonic superfluids and superconductors.

Within the present simplified model we considered a step function
momentum distribution near the Fermi surface, like in the  Fermi
gas. In reality, however, there are no reasons to expect such a
simple Fermi distribution function and there should be some smoother
distribution function near the Fermi surface even at zero temperature.
However, such a fine detail of the Fermi distribution is unimportant
for our main conclusions. Indeed, consider a probe quark well
above the Fermi surface. For this quark all details of the momentum
distribution near the Fermi surface are not important and all qualitative
outputs of the gap  and the Bethe-Salpeter
equations with the step-function distribution will
be valid.

\begin{figure}
\includegraphics[width=0.8\hsize,clip=]{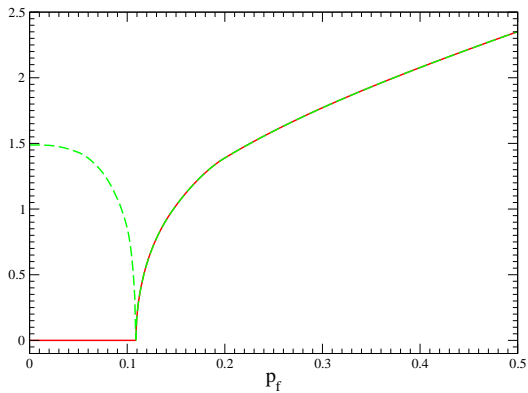}
\caption{Masses of the pseudoscalar $0^{-+}$ (solid) and scalar 
$0^{++}$ (dashed) mesons 
in units of $\sqrt \sigma$ as functions of the Fermi momentum
in the same units.The chiral restoration point is at $p_f^{cr} = 0.109$.}
\end{figure}

The other critical question is to which extent our main conclusion depends
on our specific model. Actually for a presence of the gap in the quarkyonic
matter one needs only some very general ingredients. The system must
be confined and it certainly follows from the large $N_c$ arguments
for pressure. At the same time the chiral symmetry must be restored
at some critical chemical potential. 
Then our conclusions are quite general and do not rely specifically
on the model. This model suggests an insight and a possible microscopic
scenario. 

{\bf 3. Conclusions.}
 As a conclusion we have demonstrated
that above the chiral restoration point the quarkyonic matter
represents a Fermi system with a gap. 
This gap increases with the chemical potential and is arbitrary large.
Then there should be some
processes without dissipation of energy, like in the superfluid or
superconducting systems. A mechanism of the gap formation in the
quarkyonic matter is essentially different as compared to the standard
fermionic systems. In the latter it proceeds via the condensation
of the quasibosonic Cooper pairs. In our case there is no such a
condensation and the mechanism is based on the simultaneous
properties of confinement and manifest chiral symmetry of hadrons
on top of the Fermi sea above the critical density.

We have demonstrated this property in the case of the large $N_c$
world. Then such an unusual fermionic system exists at least in
the large $N_c$ limit and consequently represents a scientifically
well defined system and it is a valid and well formulated question
to study properties of such a Fermi system. 
Needless to mention that the same result will persist for the
't Hooft model, with the only difference that in that case
there is no spin for mesons and the only quantum numbers are
parity, isospin and the radial quantum number. Contrary to the
present model the t'Hooft model is real QCD, albeit in two 
dimensions.

How will the properties of the
quarkyonic matter  differ at $N_c=3$? At present we cannot
answer this question. But it is difficult to imagine that there
could be a dramatic difference in this respect 
between the real world and the large
$N_c$ world. Indeed,  properties of QCD 
that are crucial for the present issue are
 confinement, chiral
symmetry and its spontaneous breaking. They are known to persist both
in the large $N_c$ world and with $N_c=3$. Then there is a reason to
believe that what we have demonstrated for the large $N_c$ world will
still be valid in the real world.

\medskip
{\bf Acknowledgements}
The author acknowledges support of the Austrian Science
Fund through the grant P19168-N16.

\end{document}